\documentclass{aastex}
\usepackage{emulateapj5,apjfonts}

 \newcommand{\kms}{km\,s$^{-1}$}
 \newcommand{\lb}{$\lambda$}

 \slugcomment{Accepted to the Astrophysical Journal}

 \shorttitle{Coronal Lines in Seyfert 1 Galaxies}
 \shortauthors{Rodr\'{\i}guez-Ardila et al.}

\begin{document}

 \title{Near-Infrared Coronal Lines in Narrow-Line Seyfert 1 Galaxies}

 \author{A. Rodr\'{\i}guez-Ardila\altaffilmark{1,2} and S. M. Viegas}
 \affil{Instituto de Astronomia, Geof\'{\i}sica e Ci\^encias Atmosf\'ericas $-$ Universidade de S\~ao Paulo,
 Rua do Mat\~ao 1226, CEP 05508-900, S\~ao Paulo, SP, Brazil}

 \author{M. G. Pastoriza}
 \affil{Departamento de Astronomia - UFRGS. Av. Bento Gon\c calves 9500,
 CEP 91501-970, Porto Alegre, RS, Brazil}
 
 \and

 \author{L. Prato\altaffilmark{1}}
 \affil{Department of Physics and Astronomy, UCLA, Los Angeles, CA
 90095-1562}

 \altaffiltext{1}{Visiting Astronomer at the Infrared Telescope facility, which
is operated by the University of Hawaii under contract from the National 
Aeronautics and Space Administration}
 \altaffiltext{2}{e-mail address: ardila@astro.iag.usp.br}

  \begin{abstract}
We report spectroscopic observations in the wavelength region 
0.8$\mu$m $-$ 2.4$\mu$m aimed at detecting near-infrared coronal lines in a 
sample of 5 narrow-line and 1 broad-line Seyfert 1 galaxies. Our 
measurements show that [\ion{Si}{6}] 1.963$\mu$m, [\ion{S}{9}] 1.252$\mu$m and 
[\ion{S}{8}] 0.991$\mu$m are present in most of the objects and are useful 
tracers of nuclear activity. Line ratios between coronal and low-ionization
forbidden lines are larger in narrow-line Seyfert 1 galaxies. A positive 
correlation between FHWM and ionization 
potential of the forbidden lines is observed. Some coronal lines 
have widths similar to that of lines emitted in
the broad line region (BLR), indicating that part of their flux originates 
in gas close to the outer portions of the BLR. Most coronal lines are 
blueshifted relative to the systemic velocity of the galaxy and this shift
increases with the increase in line width. Assymetries towards the blue are
observed in the profiles of high-ionization Fe lines, suggesting that the
emitting gas is related to winds or outflows, most probably originating 
in material that is being evaporated from the torus. This 
scenario is supported by models that combine the effects of shock ionization 
and photoionization by a central continuum source in the gas clouds. 
The agreement between the coronal line emission predicted by the models 
and the observations is satisfactory; the models reproduced the whole range 
of coronal line intensities observed. We also report the detection of 
[\ion{Fe}{13}] 1.074,1.079$\mu$m in three of our objects and the first 
detection of [\ion{P}{2}] 1.188$\mu$m and [\ion{Ni}{2}] 1.191$\mu$m in a 
Seyfert 1 galaxy, ARK\,564. Using the ratio [\ion{P}{2}]/[\ion{Fe}{2}] we 
deduced that most Fe present in the outer NLR of ARK\,564 is locked up 
in grains, and the influence of shocks is negligible.
  \end{abstract}

 
\section{Introduction}
Coronal lines are collisionally excited forbidden
transitions within low-lying levels of highly ionized 
species ($\chi \geq$ 100 eV). They can be formed either in 
gas photoionized by a hard UV continuum \citep{gr78, kf89, fkf97}
or in a very hot, collisionally ionized plasma
\citep{vac89}. It is also possible
that the excitation mechanism is a mixture of collisional
ionization at the shock front and photoionization in
the emitting clouds \citep{cv92, vc94, cpv98}.
In active galactic nuclei (AGN), the presence of
coronal lines, mainly in the optical region (i.e. [\ion{Fe}{7}] and 
[\ion{Fe}{10}]), have long been known to be common features in 
the spectra of these sources, although the physical conditions 
of the gas from which they originate and the location of the 
emitting region are still poorly determined.

Observationally, coronal lines are, on average, broader and 
blueshifted, relative to the centroid position, of lines of 
lower ionization stages such as [\ion{O}{3}]\lb5007 
\citep{dro84, dro86, vei91, eaw97}. This has led to the 
speculation that they are formed in a separate region,
known as the coronal line region (CLR), located at an
intermediate distance between the classical narrow line 
region (NLR) and the broad-line region (BLR). Variability
studies on [\ion{Fe}{7}] \lb6087 and [\ion{Fe}{10}] \lb6374
carried out by \citet{veill88} also suggest that the
most likely place for the CLR is between the BLR and NLR.
\citet{eaw97} found that the coronal 
line emission occurs predominantly in objects with a soft 
X-ray excess and suggested a relationship between these
lines and the X-ray absorption edges (also known as warm 
absorbers) seen in 50\% of AGN. In fact, \citet{por99} 
demonstrated through photoionization modeling 
that the optical coronal lines could be formed in the 
warm absorber and that they may strongly constrain the physical 
parameters of that medium. However, the
number of coronal lines effectively detected in the optical
region is small ([\ion{Fe}{7}]
\lb5721, \lb6087; [\ion{Fe}{10}] \lb6374; [\ion{Fe}{11}]
\lb7892 and [\ion{Fe}{14}] \lb5303). It is then necessary to
expand the range of important diagnostics of the physical 
conditions that prevail in the coronal gas to other 
wavelength intervals. In this respect, the near-infrared (NIR) is
promising because it offers a wealth of highly ionized
species different from Fe.  

Uptil now, coronal line emission between 1.5$\mu$m and 4$\mu$m  
has been detected in only a small number of AGN. 
\citet{om90} reported, for the first time,
the observation of [\ion{Si}{6}] 1.962$\mu$m and [\ion{Si}{7}]
2.483$\mu$m in the archetypical Seyfert 2 galaxy NGC\,1068. 
\citet{oli94} detected, in addition to those two lines, 
[\ion{S}{9}] 1.252$\mu$m, [\ion{Ca}{8}] 2.321$\mu$m and 
[\ion{Si}{9}] 3.9346$\mu$m in Circinus, another nearby Seyfert 
2 nucleus. \citet{mar94} detected [\ion{Si}{6}] 1.962$\mu$m 
in eight AGN (including the former two objects) in a sample
composed of Seyferts, Starburst and Ultraluminous Infrared
Galaxies (ULIRGs). They concluded that [\ion{Si}{6}] emission was 
a common characteristics of Seyfert nuclei, consistent with modeling
the line formation by photoionization of the active
nucleus. That result had reinforced the use of 
[\ion{Si}{6}] 1.962$\mu$m as a diagnostic of AGN activity
and had allowed the detection of hidden AGN, mostly
in ULIRGs samples \citep{vsk99, msm00}. \citet{tho95} studied 
NGC\,4151 in detail and reported the detection of several
coronal lines in the 0.87$\mu$m $-$ 2.5$\mu$m interval,
most notably [\ion{S}{8}] 0.911$\mu$m, [\ion{Fe}{13}] 1.074,1.079$\mu$m,
[\ion{S}{9}] 1.252$\mu$m, [\ion{Si}{6}] 1.963$\mu$m and
[\ion{Si}{7}] 2.481$\mu$m. 
\citet{grr95} studied a sample of
six Seyfert 1 galaxies (one of them NGC\,4151) in 
the 2$\mu$m region and detected [\ion{Si}{6}] 1.963$\mu$m
in all the galaxies and [\ion{Ca}{8}] 2.321$\mu$m in two
of them. 

In this paper we present the results of a search for
coronal emission lines in the wavelength range 
0.8$\mu$m $-$ 2.5$\mu$m by means of long-slit 
spectroscopy at moderate resolution ($R \sim$ 750).
The sample chosen for this study is composed of 
six Seyfert 1 galaxies, five of them classified 
as narrow-line Seyfert 1 (NLS1).
These data are the first measurements
of coronal lines in the NIR made on this sub-class of 
objects. Here, we concentrate on the analysis of the
line ratios derived from our measurements and the
physical conditions for the CLR that they imply. We 
also discuss the kinematics of the
CLR based on the analysis of the line profiles. In 
addition, we report the detection of other forbidden lines 
and molecular lines emitted by the NLR in the NIR, 
which are also useful for constraining the various excitation 
models proposed so far. 

Our observations are described in $\S$~\ref{obs}, the main
results in $\S$~\ref{res}, the study of the kinematics 
of the CLR in $\S$~\ref{kin} and the physical conditions
for the CLR derived from the observations in 
$\S$~\ref{physical}. Some comments about the low ionization
and coronal lines observed appear in$\S$~\ref{low}. We
present the main conclusions 
of this work in $\S$~\ref{final}.

\section{Observations and data reduction} \label{obs}

NIR spectra providing continuous coverage
between 0.8 $\mu$m and 2.4 $\mu$m were obtained at the
NASA 3\,m Infrared Telescope Facility (IRTF) on 2000 October
11 (UT) with the SpeX facility spectrometer \citep{ray01}. 
The detector consists of a
1024$\times$1024 ALADDIN 3 InSb array with a spatial scale
of 0.15$\arcsec$/pixel. Simultaneous wavelength coverage
was obtained by means of prism cross-dispersers. A
0.8$\arcsec \times$15$\arcsec$ slit was used during the
observations, giving a spectral resolution of 320 km/s.
The seeing was near 1$\arcsec$ during the exposures. 

Details of the spectral extraction and wavelength
calibration procedures are given in \citet{ra02a}. 
The spectral resolution provided by SpeX was 
sufficiently high so that except for a few particular
cases, the lines emitted by the NLR were spectroscopically 
resolved.

Table~\ref{tab1} lists the objects observed and
their main characteristics. There are several reasons for the
predominance of NLS1 in the sample.
These objects usually show stronger and more 
luminous optical coronal lines than other AGNs \citep{rapd00}. 
In the X-ray, NLS1s
outnumber other AGN in regard to the presence of soft
X-ray excesses, allowing to study the link
between X-rays and and coronal line intensity. NLS1 emission lines, even
those formed in the BLR, are sufficiently narrow
to allow an easier deblending of adjacent 
emission features. These characteristics make NLS1s
optimal targets for studying coronal lines and their
relationship with the NLR and BLR.

The calibrated spectra were corrected for Galactic extinction, 
as determined from the 
{\it COBE/IRAS} infrared maps of \citet{shl98}.
The value of the Galactic {\it E(B-V)} is listed in 
Table~\ref{tab1}. Finally, each spectrum
was shifted to rest wavelength. The value of $z$ adopted
was determined by averaging the redshift measured from the 
strongest lines, usually \ion{O}{1} \lb8446, [\ion{S}{3}]
\lb9531, \ion{Fe}{2} \lb9997, Pa$\delta$, \ion{He}{1} \lb10830,
\ion{O}{1} \lb11287, Pa$\beta$, Pa$\alpha$ and Br$\gamma$.
In all cases, the radial velocities were in very good 
agreement with the values reported in the literature.

In addition to the NIR data, optical spectra for 1H\,1934-063,
MRK\,335, TON\,S\,180 and MRK\,1044 were available in the
wavelength region that includes the [\ion{Fe}{7}] \lb5721
and [\ion{Fe}{10}] \lb6374 lines. There is also
data covering the [\ion{Fe}{11}] \lb7894 line for 1H\,1934-063. Details
of these observations and the reduction procedure are described 
elsewhere \citep{rapd00, ra02b}. Before making any measurements
using the optical spectra, they were also corrected for 
Galactic extinction (see values in Table~\ref{tab1}).
These data will be used 
along with the NIR spectra to study the kinematics and physical
conditions of the CLR.

\section{Results} \label{res}

Figures~\ref{hcor} to~\ref{tonscor} show the most important 
NIR narrow lines,
including the coronal lines,
observed in the galaxies. The measured line fluxes are
summarized in Table~\ref{fluxes}. The errors quoted
reflect solely the 2$\sigma$ uncertainty in the placement of the 
continuum and in the S/N around the line of interest. 
Fluxes were measured by fitting Gaussians to each emission 
feature and the continuum underlying each line was 
approximated by a low-order polynomium. In all cases, 
the NLR lines were well described
by a single Gaussian component. Measurements of [\ion{Si}{6}]
1.963$\mu$m and [\ion{Ca}{8}] 2.321$\mu$m were complicated by
poor atmospheric transmission near the former
and the presence of deep CO $\Delta\nu =2$ bands in the
latter. However, the residuals around the features of interest, 
after dividing the target spectra by that of the telluric star, are 
low enough in  most objects to allow an unambiguous 
identification of the lines.

In addition to the coronal lines, Table~\ref{fluxes} also list the fluxes
of other forbidden and permitted lines detected in the NIR spectra.
We found that [\ion{S}{3}] 0.953$\mu$m is, by far, the strongest 
of all narrow forbidden lines and one of the brightest lines in the
NIR range, particularly in 1H\,1934-063 and 
ARK\,564. Molecular H$_{2}$ emission is also observed in these
two objects as well as in the broad-line Seyfert 1 
galaxy NGC\,863. 

[\ion{Fe}{13}] 1.074$\mu$m is observed unambiguously in
1H\,1934-063 (Fig.~\ref{hcor}), ARK\,564 (Fig.~\ref{arkcor}), 
MRK\,335 (Fig.~\ref{335cor}) and  TON\,S\,180.
That line has only been reported before in
NGC\,4151 \citep{tho95} and NGC\,1068 \citep{oli01}.
Evidence of the companion [\ion{Fe}{13}] 1.079$\mu$m line is
seen in 1H\,1934-063 and ARK\,564 but it is too blended
with \ion{He}{1} 1.083$\mu$m to allow a confident measurement of
its flux. [\ion{P}{2}] 1.188$\mu$m and [\ion{Ni}{2}] 1.191$\mu$m
are observed in ARK\,564 (see Figure~\ref{arkcor}). The only 
previous identification of
[\ion{P}{2}] in an extragalactic object was in 
NGC\,1068 \citep{oli01}. [\ion{Ni}{2}], to our
knowledge, was never reported in an extragalactic source. Its
presence indicates gas with densities below 500 cm$^{-3}$,
which is the critical density of that line.  

Our data show that not all AGN display coronal lines. In fact, except
for a few lines, they
are not detected above the noise level in MRK\,1044 and
TONS\,180. In this latter object, the only coronal line detected
was [\ion{Fe}{13}]. In contrast, 1H\,1934-063 and Ark\,564
display very rich CLR spectra, with bright 
[\ion{Si}{6}] 1.963$\mu$m, [\ion{Si}{10}] 1.430$\mu$m, 
[\ion{S}{9}] 1.252$\mu$m, [\ion{S}{8}] 0.991$\mu$m
 and detectable [\ion{Ca}{8}]
2.321$\mu$m. MRK\,335 seems to be an 
intermediate case, with detectable emission at
the expected position of most lines. 

The above findings agree with those of \citet{mar94}, 
who conducted a survey of [\ion{Si}{6}] 1.962$\mu$m
in 26 galaxies, 15 of them classified as Seyferts. 
They detected this line in some but not all of the 
AGN in their sample. They interpreted their results 
as indicating that either the conditions
required to produce bright coronal lines are not 
always met or the 2$\mu$m extinction toward the CLR 
is too large in the objects in which it was
not observed.    

It is important to compare the similarity of 
the NLR spectra of the different objects of our 
sample in the NIR 
region. For this purpose, line ratios
between  coronal and forbidden low-ionization
lines were calculated, as is shown in Table~\ref{corrat}.
Compared to data published in the literature for the
Seyfert 2 galaxies NGC\,1068 and Circinus and the broad-line
Seyfert 1 NGC\,4151, some line ratios in the NLS1 exhibit  
different behavior. [\ion{S}{9}] 1.252$\mu$m/[\ion{Fe}{2}] 1.257$\mu$m,
for instance, is six times stronger in ARK\,564
than in Circinus and 12 times larger than in NGC\,4151. 
[\ion{S}{8}] 0.991$\mu$m/[\ion{Ca}{1}]
0.985$\mu$m is larger in ARK\,564 than in Circinus and 
NGC\,1068, by factors of 5 and 12 respectively. 
This trend is also observed 
in line ratios involving coronal and molecular lines,
for example as in [\ion{Ca}{8}] 2.321$\mu$m/H$_{2}$ 2.121$\mu$m.
This ratio is up to four times higher in 1H\,1934-063 than in
Circinus. Although it is not possible to draw any definitive
conclusions on this matter because of the small number of 
objects, there is the tendency for NLS1 to display extreme 
values of a given ratio, even compared to broad-line
Seyfert 1 objects. Whatever the case, it cannot be the result
of reddening because every line pair involved in the ratios listed in 
Table~\ref{corrat} are so close in wavelength that they are
free from this effect.

Table~\ref{opfluxes} lists the line fluxes of the 
coronal lines measured from the optical spectra. For ARK\,564
the data were taken directly from \citet{eaw97}. 
Although the slit width used in the
optical spectroscopy was larger than in the NIR
(2$\arcsec$ and 0.8$\arcsec$, respectively), it has
little effect on the high excitation lines when combining
data from these two spectral regions. A careful
inspection of the 2-D frames at the expected position
of the coronal lines does not reveal any sign  of extended 
emission. Spatially, the CLR emission is indistinguishable 
from the unresolved nuclear emission. Similar results 
are reported for NGC\,1068 \citep{mar96} and Circinus
\citep{oli94}. Nonetheless, we recall that the galaxies 
in our sample are located at relatively large distances 
(42 $\leq$ D $\leq$ 148 Mpc), meaning that an upper limit to 
the size of the CLR based on the FWHM of the unresolved 
nucleus (in the 0.8$\arcsec$ resolution case) would vary from 
150 pc in 1H\,1934-063  
to 950 pc in TONS\,180. For the latter object, that size is
large enough to include the extended narrow line regio, but only
one coronal line was detected. For the former, the CLR is 
well within the NLR limits. Additional constraints on the
location of the CLR can be obtained by the study of the line
profiles and the kinematics they imply.

\section{Kinematics of the CLR} \label{kin}

The presence of correlations between the widths of 
forbidden optical lines and the ionization potential 
(IP) and critical density (N$_{\rm crit}$) of the 
corresponding transition have been reported in several 
samples of Seyfert galaxies \citep{dro84, dro86, ev88}. 
This result was interpreted in terms of a stratified NLR
in which high ionization lines (i.e. coronal lines) are
emitted in the inner NLR or even in the transition region between
the NLR and BLR. Lower ionization lines (i.e. [\ion{O}{1}],
[\ion{S}{2}], and [\ion{N}{2}]) originate in the
outer portions of the NLR. 

Nonetheless, according to the published literature, 
when NIR coronal lines are involved, the
correlations are not as clear as those observed with 
optical lines. \citet{grr95} studied a sample of six 
Seyfert 1 galaxies in the
2$\mu$m region. They detected [\ion{Si}{6}] 1.96$\mu$m
in all the galaxies and found that its width
varied from very narrow profiles ($\sim$300 \kms) 
to profiles approaching that of the broad lines ($\sim$2200 \kms).
[\ion{Ca}{8}] 2.32 $\mu$m, detected in two objects of
Giannuzzo et al.'s sample, showed no correlation with 
the [\ion{Si}{6}] width, even though they have IPs of the
same order (127 eV and 167 eV, respectively). 
\citet{kno96}, in their spatially
resolved 1.24$-$1.30 $\mu$m spectroscopy of
NGC\,4151, found [\ion{S}{9}] 1.252$\mu$m to be narrower
than Pa$\beta$ (narrow component) and almost 2.5 times 
narrower than the low$-$ionization line 
[\ion{Fe}{2}] 1.2567 $\mu$m, for which they
measured a FWHM of 400 \kms. Combining Giannuzzo et al.'s
data with that of \citet{kno96} for NGC\,4151, no correlation
between IP and coronal line width is observed.

Similar results have been reported when NIR lines of
Seyfert 2 galaxies are studied. In NGC\,1068 \citep{mar96}
no trend between IP and FWHM is observed. Low and
high ionization lines have similar FWHM ($\sim$ 1000 \kms). 
\citet{oli94}, in their study of the Circinus 
galaxy, reported that all the visible and IR coronal 
lines were quite narrow and only barely resolved, with 
intrinsic widths of less than 100 \kms. 

Table~\ref{fwhm} lists the FWHM, already corrected for
instrumental width, and the shift of the line centroid with
respect to the systemic velocity of the galaxy, for the
forbidden lines reported in Table~\ref{fluxes}.
MRK\,1044 and TON\,S\,180 were not included because they
lack significant coronal line emission.
Also listed in Table~\ref{fwhm} are the FWHM and shifts measured 
for the optical coronal lines [\ion{Fe}{7}], 
[\ion{Fe}{10}] and [\ion{Fe}{11}] measured in 1H\,1934-063,
ARK\,564 and MRK\,335. 

\subsection{Correlations between line width, ionization potential and critical density}

Several interesting results arise from the data
listed in Table~\ref{fwhm}.
In 1H\,1934-063, ARK\,564 and MRK\,335, there is
a clear correlation between line width and IP,
as can be graphically seen in Figure~\ref{ipvsfwhm}. No
correlation is observed for NGC\,863. 
This fact can be explained if we recall that most NLR lines 
in that object were unresolved and/or
detected only as upper limits. Except for [\ion{S}{8}] 0.991 $\mu$m
(IP=280 eV), the lines of NGC\,863 plotted in Figure~\ref{ipvsfwhm}
have FWHM equal to the instrumental value. 
Overall, low-ionization 
lines (i.e. [\ion{Ca}{1}], [\ion{Fe}{2}]) are characterized 
by values of FWHM lower than those of the
highest ionized species (i.e. [\ion{Fe}{13}]
and [\ion{Si}{10}]). In 1H\,1934-063, for example, there is
up to a factor of four difference in FHWM between low and high
ionization lines. Figure~\ref{ipvsfwhm}
shows the presence of a zone of avoidance in width for 
lines of small IP. No low ionization lines are found in the
upper left corner of the plot, the region of large FWHM and
small IP. Assuming that the width of the lines reflects the 
bulk motion of the emitting clouds in the gravitational
potential of the central mass concentration, our results 
imply that low ionization lines are preferentially emitted 
in the outermost portion of the NLR. 
Among coronal lines, we
notice a spread in line width, the most noticeable being
the case of [\ion{S}{9}], which is as narrow as [\ion{Fe}{2}].
However, this fact does not contradict our results,
mainly because we found evidence that ions of different
species follow a different slope for the FWHM-IP correlation.

In effect, the left panels of Figure~\ref{ipvsfwhm1} show that for
1H\,1934-063, ARK564 and MRK\,335, the sources with the most 
conspicuous CLR spectrum, the width of the iron lines increases
faster with the IP than that of sulfur and
silicon. [\ion{Si}{10}] 1.430$\mu$m, the line with the largest
IP (351 eV), is not the broadest one. [\ion{Fe}{13}] 1.074$\mu$m, 
with a slightly smaller IP 
(331 eV), is significantly broader than [\ion{Si}{10}].
[\ion{S}{9}] 1.252$\mu$m (IP=328 eV), another high-ionization
coronal line, has a width comparable to those of low-ionization 
lines (nonetheless, this line was not detected in MRK\,335). 
In 1H\,1934-063, there is a slight decrease in line
width between [\ion{Fe}{10}] and [\ion{Fe}{11}], not 
observed in ARK\,564 and MRK\,335. But clearly, when the
degree of ionization of the iron lines increases, the trend is to
increase the line width, also.

A similar tendency is observed when the gas kinematics 
are plotted against the critical density (N$_{\rm crit}$) of the
different transitions, as can be seen in the right panels of
Figure~\ref{ipvsfwhm1}. For 1H\,1934-063, we observe  that
the gas emitting the sulfur lines has different kinematics
than that emitting the iron or silicon lines, as previously
suggested. As in the
case of the ionization potential, the width of the sulfur lines
is insensitive to the increase of the critical density. For
ARK\,564 and MRK\,335, all lines of different species
tend to positively correlate with N$_{\rm crit}$, 
reinforcing the hypothesis of a stratified CLR. Nonetheless,
it should be kept in mind that the critical densities of high
ionization NIR lines are lower than those of the optical
region, even though the degree of ionization of a particular
ion increases. This fact may create an artificially weaker 
correlation between FWHM and N$_{\rm crit}$ compared to
that observed between FWHM and IP. 

\subsection{Evidences for outflows in the CLR}

Why do different species not follow a similar FWHM $-$ IP 
or FWHM $-$ N$_{\rm crit}$ correlation? Clues to the answer
can be obtained by examining the values of $\Delta$V exhibited by the 
coronal lines (see Table~\ref{fwhm} and Figure~\ref{ipvsfwhm1}). 
While low-ionization lines have no shifts
at all or are redshifted by a small amount, coronal lines 
are characterized by large blueshifts, the largest 
ones being for the lines with the largest IP ($\geq$ 300 eV) and 
FWHM. Both in 1H\,1934-063 and ARK\,564, the increase/decrease 
in blueshift clearly accompanies the increase/decrease in line 
width and IP for lines of the same species: 
[\ion{Fe}{13}] is broader and more blueshifted than 
[\ion{Fe}{7}]. Interestingly, and most importantly,
[\ion{Fe}{13}], the broadest 
coronal line, is the most blueshifted line in these two objects. 
[\ion{Si}{10}] and [\ion{S}{9}], with IP similar to that 
of the former, but significantly narrower, are less 
blueshifted. In 1H\,1934-063, the decrease in line width
when going from [\ion{Fe}{10}] to [\ion{Fe}{11}] is also
accompanied by a decrease in blueshift. 
MRK\,335 behaves somewhat differently. While [\ion{Fe}{13}] is again 
the broadest coronal line, the most blueshifted one 
is [\ion{Si}{10}]. In addition, iron lines of different IP seem to be 
characterized by a similar shift. It is also important to note
that in this object the
sulfur and silicon gases have rather similar kinematics
but, as in the other two objects,  different from that of iron. 
For the remaining three galaxies 
(NGC\,863, MRK\,1044 and TON\,S\,180), nothing can be said about
the kinematics of the CLR due to the lack of sufficient 
information. Nonetheless, it seems that when
significant coronal line emission is detected, the
iron lines manifest different kinematics than the rest of
the coronal lines. Although at the present time
it is not possible to establish how significant is the result
found for Fe because of the small number of objects,
we consider that the results are highly encouraging and lead
towards a breakthrough in the current understanding of the CLR.

If the blueshift of the lines is interpreted as being attributable
to an outflow of coronal gas towards the observer, what is the
approximate location of such a flow? Figure~\ref{ipvsfwhm1}
is useful for addressing this question. There, we see that
the FWHM of [\ion{Fe}{13}] and [\ion{Si}{10}] are
larger than that of \ion{O}{1} 1.128$\mu$m. 
Permitted \ion{O}{1} emission is a feature
completely associated with the BLR \citep{gr80, mw89}. 
Very recently, \citet{ra02b} 
presented observational evidence that this line arises in
the outer boundary of that region. Therefore, our data 
suggest that part of the CLR is located in the boundary 
between the BLR and NLR. At this location, a natural
scenario for the origin of the flow is, for example, 
radiatively accelerated material evaporated from the outer 
regions of the accretion disk and/or inner walls of the 
torus, as suggested by \citet{eaw97} for ARK 564 
(and other objects). \citet{pv95} had already proposed 
a similar scenario in which molecular clouds located at the inner
edge of the torus are heated, ionized and evaporated by
intense UV/X-ray radiation from the AGN.  Another possibility is 
interaction between the radio jet and ambient gas. Nonetheless, the 
NLS1 of our sample are radio-quiet objects. In addition, studies in the
optical region aimed at detecting coronal lines in radio-loud
AGN \citep{ea94, eaw97} have found that these objects lack
significant coronal line emission, giving little
support to this latter hypothesis.

\subsection{Ionization structure of the CLR}

Based on the results from the above section, we propose
a CLR in which high ionization iron and silicon lines 
are formed in the bordeline between the BLR and NLR, 
the bulk of the [\ion{Fe}{13}] line being emitted closest to the torus. 
Farther out, but yet in the BLR$-$NLR interface, [\ion{Fe}{10}], 
[\ion{Fe}{11}] and [\ion{Si}{10}] are formed. The remaining 
forbidden lines would be emitted in the classical NLR. A good 
test for this scenario is the analysis of a sample composed
of Seyfert 1 and 2 galaxies. It is expected that the latter
objects show little or no [\ion{Fe}{13}], as is already the case for
NGC\,1068 and Circinus.  Also consistent with this scenario is
the lack of correlation between the FWHM and IP and N$_{\rm crit}$
of high-ionization coronal lines in Seyfert 2s. The lines
emitted closest the torus are partially hidden from direct
view under certain orientation angles. For Seyfert 1s, if
the conditions for forming coronal lines are met, they are 
observed in all their extent. 

\section{Physical properties of the CLR} \label{physical}

Based on the results of the above section, important clues
about the physical conditions and excitation mechanisms of 
the coronal gas in Seyfert galaxies can be drawn. As 
Table~\ref{corrat} shows, the NLR of the objects is 
characterized by a mixture of intense low and high ionization 
lines. While the latter lines are blueshifted
relative to the systemic
velocity of the host galaxy, the former are, in
general, coincident with the rest position of the line. 

The above means that at least two different components
are needed in order to explain the observed NLR spectrum:
one set of clouds producing low to intermediate ionization
lines (i.e. [\ion{Ca}{1}], [\ion{S}{3}]) and another set,
closer to central source, and responsible for coronal
lines (i.e. [\ion{Si}{6}], [\ion{Fe}{10}], [\ion{S}{8}]). 
In this section, based on current photoionization models,
we make rough estimates of the necessary
conditions responsible for the emission of coronal lines.
A detail modeling of the whole NLR spectrum for the
individual objects (ARK\,564, 1H\,1934-063 and MRK\,335) 
is left for a future paper. 

Table~\ref{opnir} lists ratios
between optical and NIR coronal lines, found in 
1H\,1934-063 and ARK\,564. The values listed are corrected 
for internal reddening, derived from the Balmer decrement
assuming an intrinsic ratio of 3.1. 
For the former galaxy, the E(B-V) was
determined by \citet{rapd00} while
for the latter, it was taken from \citet{eaw97}. We
have chosen these two objects because all coronal lines
between 0.5$\mu$m and 2.5$\mu$m predicted by models are
observed in these two sources. For comparison, data 
for NGC\,1068 and Circinus (also reddening corrected) 
are included.

It is interesting to see that the four AGN have similar 
values of most line ratios, differing by 
less than a factor of 3. The exception is
NGC\,1068, which presents extreme values in the ratios 
[\ion{Fe}{10}]/[\ion{Fe}{7}] and [\ion{S}{9}]/[\ion{Si}{6}].
This was already noted by \citet{mar96} in their
analysis of the NIR and visible coronal lines of that object.
They found that NGC\,1068 has a somewhat lower ionization
CLR, which they attributed to either a steeper ionizing
continuum or a somewhat lower ionization parameter.

Assuming that the coronal lines are formed by the combination
of photoionizing radiation from the active center and a
shock front (from the evidence discussed in the previous
section), we have compared the observed coronal line ratios with 
the grid of model predictions of \citet{cv01}. The 
simulations apply whether the shocks originate from a
radial outflow of clouds (if evaporated material from the
torus give rise to the coronal lines) or from an interaction
of the emitting clouds with a radio jet.

Table~\ref{opnir} shows the predictions of the different
models obtained by varying the shock velocity and the
preshock densities, keeping constant the geometrical
thickness of the clouds and the intensity of the 
photoionizing radiation of the central source.
Also listed in Table~\ref{opnir} are the predictions
of the best fitting model obtained for Circinus by
\citet{cpv98}. It
was obtained from a suitable combination of 
clouds, taking into account the effect of geometrical
dilution and a weighted average of single cloud
spectra showing different characteristics (varying
preshock velocity and density).

Clearly, the model predictions of Table~\ref{opnir} 
cover the range of line ratios found in the objects of 
our sample, as can be seen in Figure~\ref{obsvsmod}, 
and offer a suitable scenario to explain
the origin of the coronal lines in AGNs. The advantage
of this approach over previous ones, e.g. the 
\citet{pv95} model, is that here the
interaction between the evaporated material and the
surrounding gas is taken into account. Since not 
all clouds have the same
preshock velocity, $V_s$,  and density, $n_0$, even though 
they are subjected to the same radiation field, the 
ionization structure 
among individual clouds varies. A particular
line ratio is favored over the others according to the initial 
conditions. As an example, model 29 (M29; $V_s$=200 \kms, 
$n_0$=200 cm$^{-3}$) emits preferentially [\ion{Si}{6}],
while model 47 (M47; $V_s$=100 \kms, $n_0$=300 cm$^{-3}$)
favors the emission of [\ion{Fe}{10}].  The predominance
of a particular set of clouds leads to the enhancement  of
a given line  within the CLR. As a result, there is
a natural scatter in CLR ratios, as observed.
Under the assumed conditions the clouds
are poor emitters of intermediate and low ionization
lines. These would be produced farther out in the NLR. 

We have also compared the dereddened line ratios to
those predicted by a pure photoionization model (see last 
column of Table~\ref{opnir}), calculated
by \citet{oli94} assuming the ``standard AGN continuum''of
CLOUDY, solar abundances and $Q(\rm H)$=2$\times$10$^{52}$ 
s$^{-1}$. It can be seen that although some line ratios
are in reasonable good agreement with those observed, it is not 
possible to reproduced the whole CLR spectrum of a single
object by pure photoionization. 
This gives stronger support to the idea that the high ionization
lines arise from a set of emitting clouds having 
different physical conditions. The observed spectra
would result from the combination of the individual cloud 
spectra.

Quantitatively, the model predictions also agree
with the  observations. From Table 13 of \citet{cv01}, 
temperatures of up to 6$\times10^{5}$\,K are expected 
from downstream gas ionized by the shock models
of Table~\ref{opnir}. This temperature is of the same order of 
the value derived by \citet{eaw97} for 
the CLR in AGNs ($\sim$1.2$\times10^{5}$) using the
[\ion{Fe}{7}] lines as temperature indicators. They, in
fact, predict higher temperatures for the plasma
emitting very high ionization lines, such us [\ion{Fe}{10}]
and [\ion{Fe}{11}].  

The absence of coronal lines in some of the 
objects can be explained if the photoionizing
radiation from the central source that reaches 
the evaporated material is somewhat diminished
before reaching the torus. It is also possible
that orientation effects between the observer
and the flow shields the emission from the CLR
$-$ i.e., most clouds are redshifted and obscured
from our line of sight. This might be the
case for MRK\,1044 and TON\,S\,180, which
present a very weak CLR spectra. However, the
hydrogen column density $N_{\rm H}$ of these
objects, measured from X-ray observations, show
little or no excess of neutral H over the
Galactic value as is observed in columns 6 and 7 
of Table~\ref{tab1}).
Another possibility is that
coronal line emitting material in the line of
sight is absent, or there is a high anisotropy
in the continuum radiation that reaches
the NLR. Interestingly, low to intermediate
forbidden lines in the optical region for the above two
sources are also very weak or absent \citep{com98},
contrary to what is seen in the remaining objects.
Although the study of the reasons for the lack of 
coronal lines in some Seyferts is out of the scope 
of this paper, it is important to note that the observational
evidence presented here suggests that both low and
high ionization lines may have a common origin.
It is certainly a question that deserves to be explored. 

\section{Near-infrared Low ionization lines} \label{low}

Along with the forbidden high ionization lines, already
discussed in the above sections, forbidden low-ionization and
molecular lines are also present in the spectra. Although they
may not be fully associated with the AGN, it is instructive 
to comment about the physical processes that can give
rise to them to complement the picture already outlined
for the nuclear regions of the galaxies under consideration.
The small width observed in low-ionization and molecular
lines suggests that they are formed in the outer portions
of the NLR, where, in addition to the radiation from the central
source, thermal processes can also contribute to the observed
emission. 

\subsection{A short note about H$_{2}$ molecular lines}  

>From our  observations we note the presence
of molecular H$_{2}$ lines at 1.957$\mu$m and 2.121$\mu$m in 
ARK\,564, 1H\,1934-063 and NGC\,863.
H$_{2}$ emission is observed in a variety of
sources including infrared ultraluminous and starburst galaxies
and AGN \citep{kaw89, kw93, gvh94, gold95, vsk99}.
Several mechanisms have been proposed to excite
these lines and their emission line ratios can be used to test 
the dominant process that accounts for their observed intensity.
The H$_{2}$ ratio 2.247$\mu$m/2.121$\mu$m is commonly used to
distinguish between thermal (0.1$-$0.2) and UV excitation
($\sim$ 0.55) in low density regions. X-ray
heating can also play a role in the excitation of the molecular
gas \citep{ahe97}. 

We have not detected H$_{2}$ 2.247$\mu$m above 2$\sigma$ confidence
in any of our spectra but the observation of 2.121$\mu$m may help us 
to set up an upper limit to the intensity of the former line. If UV
excitation were important, H$_{2}$ 2.247$\mu$m should be
observed at half the intensity of 2.121$\mu$m. An inspection of 
our data shows that 2.247$\mu$m is below the required intensity. The
upper limit to its flux is I(\lb2.247)=0.2$\times10^{-15}$ erg cm$^{-2}$ s$^{-1}$
and 0.4$\times10^{-15}$ erg cm$^{-2}$ s$^{-1}$ for ARK\,564 and 1H\,1934-063,
respectively. These values imply a 2.247$\mu$m/2.121$\mu$m ratio of
0.17 and 0.40, respectively. For ARK\,564 that value suggests thermal
excitation of the H$_{2}$ lines, while for 1H\,1934-063,
it is closer the value expected for UV excitation.
It is also possible that X-ray heating plays an active
role in the formation of molecular H$_{2}$ gas. This
last possibility can be evaluated through the  
[\ion{Fe}{2}]/Br$\gamma$ ratio, which predicts values
up to $\sim$20 while in \ion{H}{2} regions that 
ratio is less than 2.5. From Table~\ref{fluxes} we
obtain ratios of 0.63 and 2.84 for ARK\,564 and 1H\,1934-063,
respectively. Those values allow us to discard 
X-ray emission as the dominant mechanism to excite the molecular
lines in these two objects and favor thermal processes,
most probably associated with star forming regions close to the AGN,
to produce the H$_{2}$ molecular emission.

\subsection{The [\ion{Fe}{2}]/[\ion{P}{2}] ratio}

The [\ion{Fe}{2}] lines in the NIR are useful tracers
of shocks in the NLR of AGN. This hypothesis
comes from the observational fact that [\ion{Fe}{2}] is
weak in \ion{H}{2} regions and planetary nebulae 
(where photoionization by a central source dominates) while
it is strong in shock-excited filaments of supernovae remnants
\citep{oli01}. The reason for the increase of 
[\ion{Fe}{2}] emission in the later sources is attributed
to the evaporation of iron-based grains by shock fronts.
It implies that the abundance of Fe$^{+}$ relative to any
non-refractory species (i.e. P$^{+}$), formed in the same 
partially ionized region, can be used to constrain the origin 
of the [\ion{Fe}{2}] emission. 

The above can be probed by means of 
the [\ion{Fe}{2}] 1.257$\mu$m/[\ion{P}{2}] 1.188$\mu$m ratio.
According to \citet{oli01}, it is large
($\geq$ 20) in regions excited by fast shocks while
low ($\leq$ 2) in normal photoionized regions. For 
ARK\,564, the only galaxy in which the [\ion{P}{2}]
1.188$\mu$m line was detected, we found that the 
[\ion{Fe}{2}]/[\ion{P}{2}] ratio is 0.82,
suggesting that shocks fronts have little or no importance 
in the production of the [\ion{Fe}{2}] emission.
\citet{oli01} also points out that [\ion{Fe}{2}]/[\ion{P}{2}]
almost solely depends on the Fe/P relative abundance.
According to their calculations, a solar Fe/P$\simeq$100 
abundance ratio implies in a [\ion{Fe}{2}]/[\ion{P}{2}] 
$\simeq$50 (i.e. assuming that all iron is in gaseous
phase), quite near to the value measured in 
supernovae remnants, where fast shocks destroy most the
grains. The low ratio found in ARK\,564 
([\ion{Fe}{2}]/[\ion{P}{2}] = 0.82) indicates that
most of the Fe is locked up into grains. We recall
that [\ion{Fe}{2}] lines are preferentially emitted
in the outer portions of the NLR, contrary to the place
where forbidden high ionization lines of this element
are formed.

\section{Summary}  \label{final}

NIR coronal lines in the 0.8$-$2.5 $\mu$m
interval are studied, for the first time, in a sample of 
Seyfert 1 galaxies. We have found that [\ion{S}{8}],
[\ion{S}{9}], [\ion{Fe}{13}], [\ion{Si}{6}] and 
[\ion{Si}{10}] are present in most of the objects.
These lines are significantly broader than 
low-ionization lines such as [\ion{S}{3}], [\ion{Ca}{1}]
and [\ion{Fe}{2}]. In addition, blueshifts of up to
550 km s$^{-1}$, relative to the systemic velocity of
the galaxies, are measured for [\ion{Fe}{13}]. The
amount of blueshift is strongly correlated with FWHM
and varies among the species with similar IP. These
results hold when NIR data are combined with 
existing optical spectroscopy. Moreover, the
FWHM of the broadest coronal lines is larger than 
that of \ion{O}{1} \lb1.128$\mu$m, a pure BLR
feature, but lower than broad Pa$\beta$. 

The above findings give strong observational support
to the picture in which coronal lines are
emitted in the intermediate region between the NLR
and BLR. The blueshifts and assymetries of the highest
ionization lines suggest that they are formed from outflow
gas, most probably associated with material evaporated from 
the torus. The combined effect of shocks between that 
material and the ambient gas and the intense radiation
field from the central source would produce the observed
coronal line emission. Models that take into account
these two mechanisms successfully reproduce the
observed values of line ratios between coronal optical
and NIR lines.

Outward from the CLR, where most low- to 
intermediate-ionization lines are being emitted,
thermal processes associated with starburst activity
may give rise to the observed molecular H$_{2}$
emission, as is indicated by the value of the H$_{2}$ ratio 
2.247$\mu$m/2.121$\mu$m. Our data also report the first
detection of [\ion{P}{2}] 1.188$\mu$m and 
[\ion{Ni}{2}] 1.191$\mu$m in a Seyfert 1 object. The
ratio between [\ion{Fe}{2}] 1.257$\mu$m and [\ion{P}{2}]
sets important constraints on the dominant excitation 
mechanism for the low-ionization gas. The low value 
found for [\ion{Fe}{2}]/[\ion{P}{2}]
(0.82) indicates that most iron is locked up in 
grains, being negligible the excitation via shocks.

  \acknowledgements

  This research has
  been supported by the Funda\c c\~ao de Amparo a Pesquisa
  do Estado de S\~ao Paulo (FAPESP) to ARA, PRONEX grant
  662175/1996-4 to SMV and ARA and PRONEX grant 7697100300 to
  MGP and ARA. The authors are grateful to an anonymous referee
  for comments and suggestions that helped to improve this article.

\centerline{\includegraphics[width=10cm]{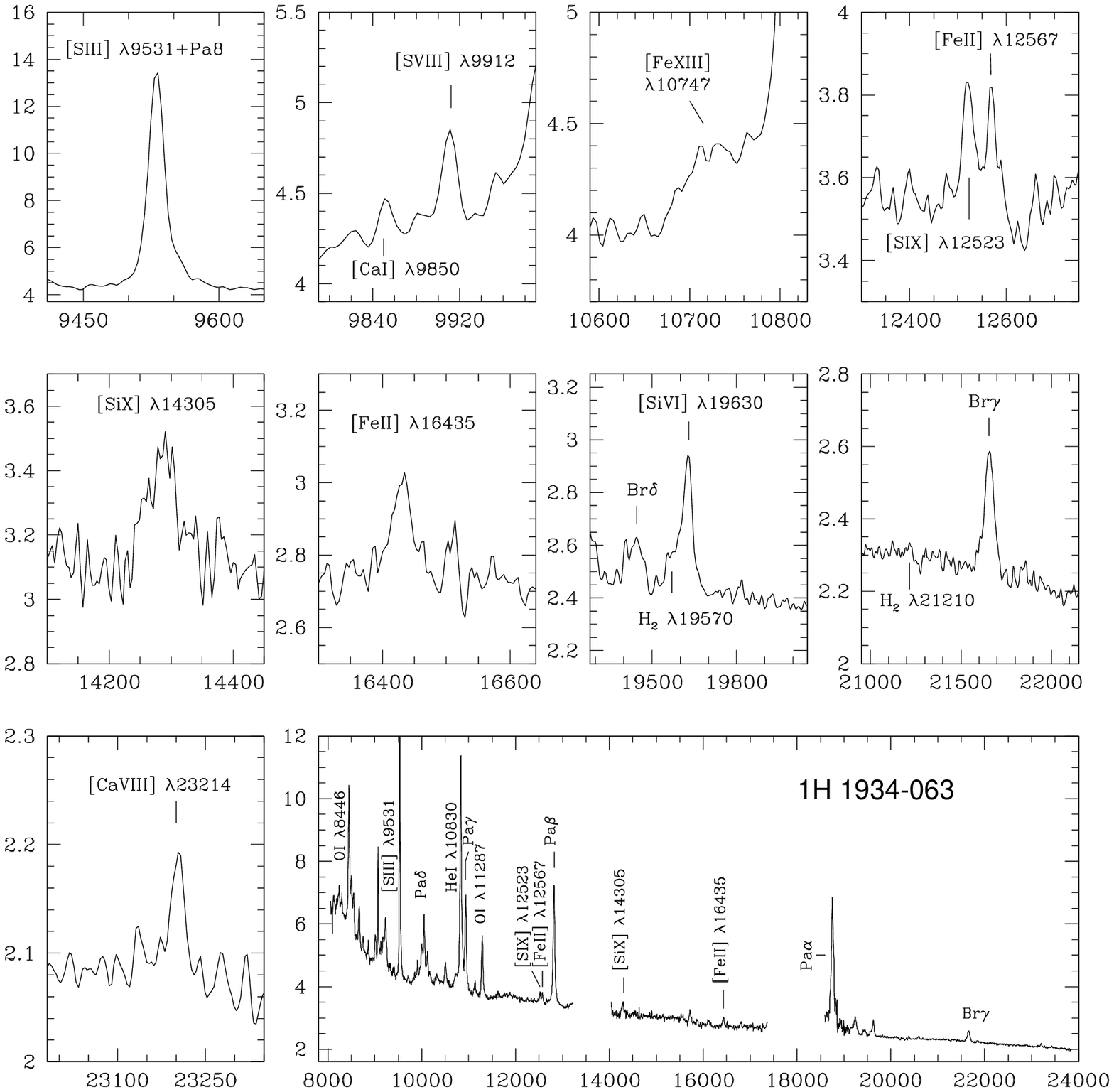}}
\figcaption{Flux calibrated spectrum of 1H\,1934-063 showing, in detail, 
the most conspicous NIR forbidden lines (in rest wavelength). The 
botton panel shows the SpeX spectrum of the source from 0.8$\mu$m to 
2.4$\mu$m with the identification of \ion{H}{1} and other permitted 
and forbidden lines. Wavelengths are in \AA\ and fluxes in units of 
10$^{-15}$ erg cm$^{-2}$ s$^{-1}$.} \label{hcor}
\centerline{}

\centerline{\includegraphics[width=10cm]{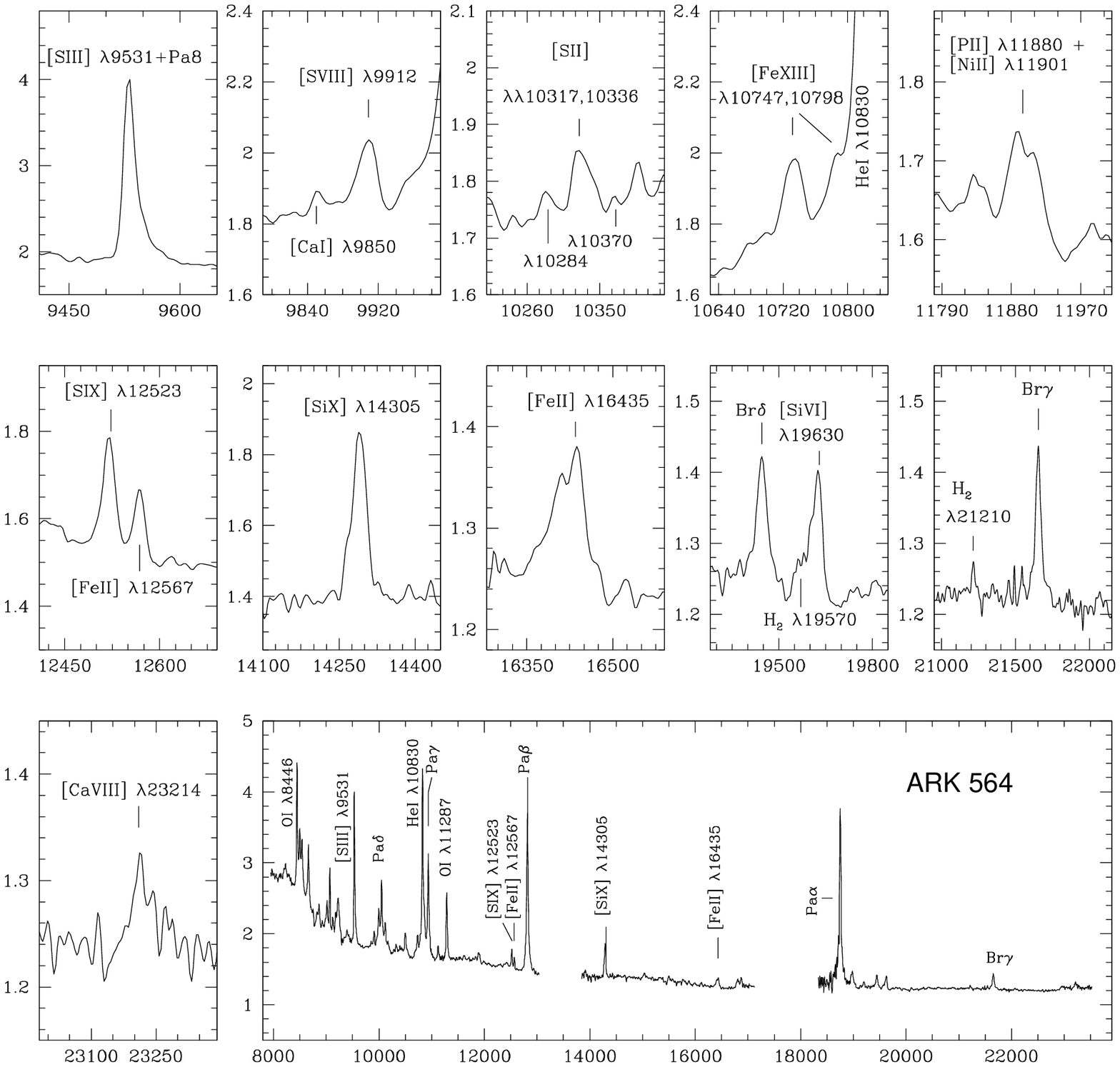}}
\figcaption{The same as Figure~\ref{hcor} for ARK\,564.} \label{arkcor}

\centerline{\includegraphics[width=10cm]{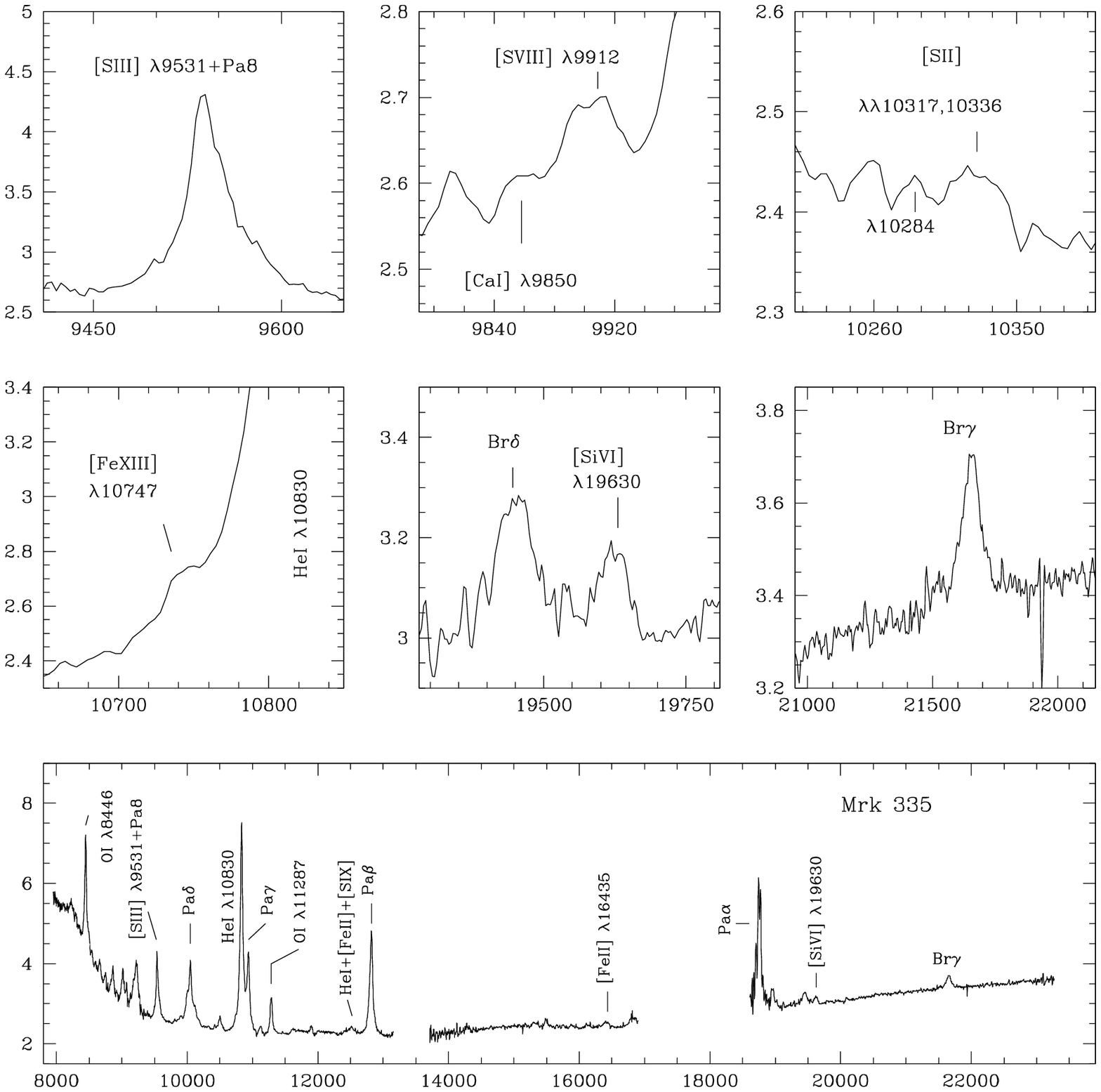}}
\figcaption{The same as Figure~\ref{hcor} for MRK\,335.} \label{335cor}

\centerline{\includegraphics[width=10cm]{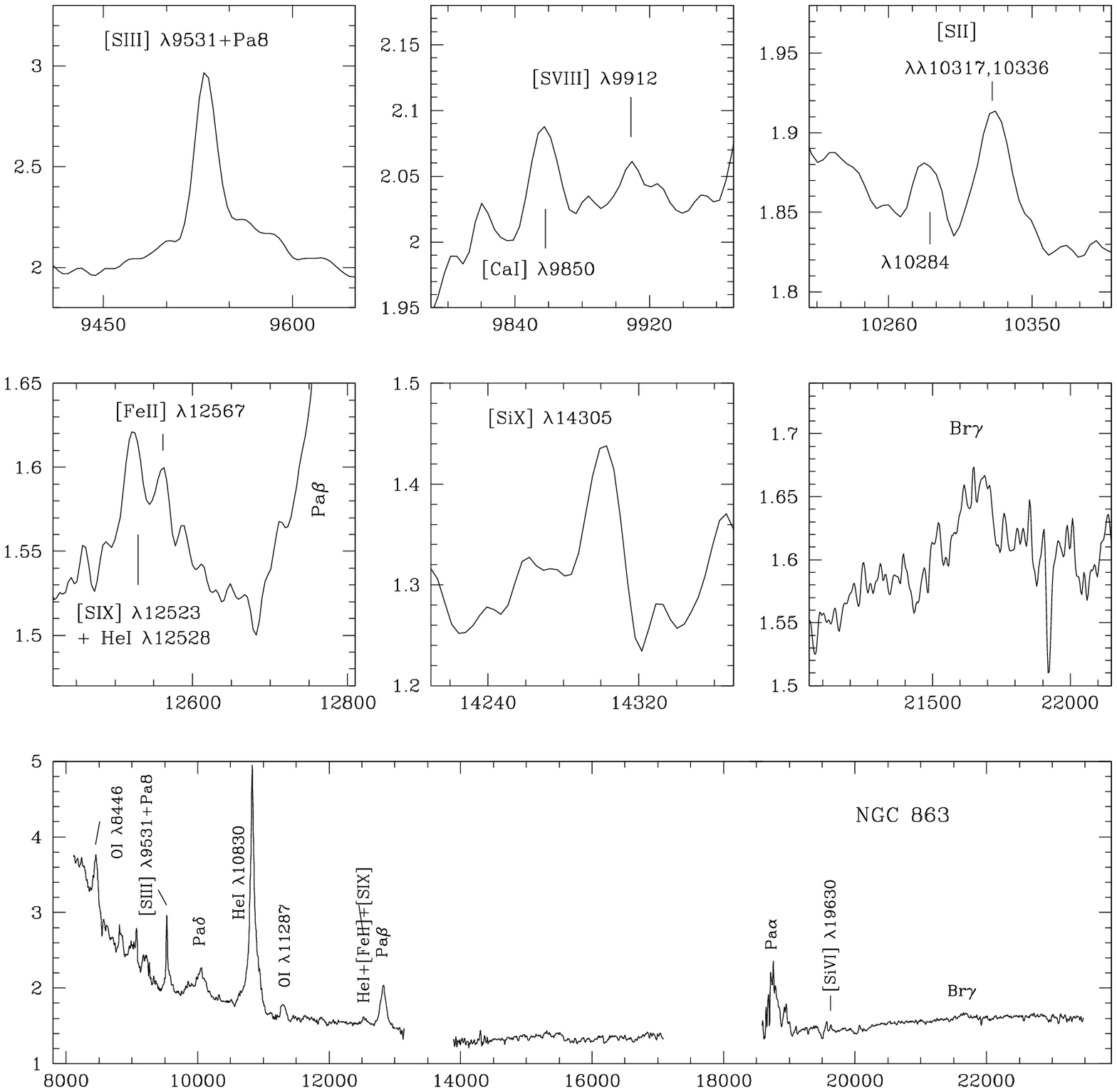}}
\figcaption{The same as Figure~\ref{hcor} for NGC\,863.} \label{ngccor}

\centerline{\includegraphics[width=10cm]{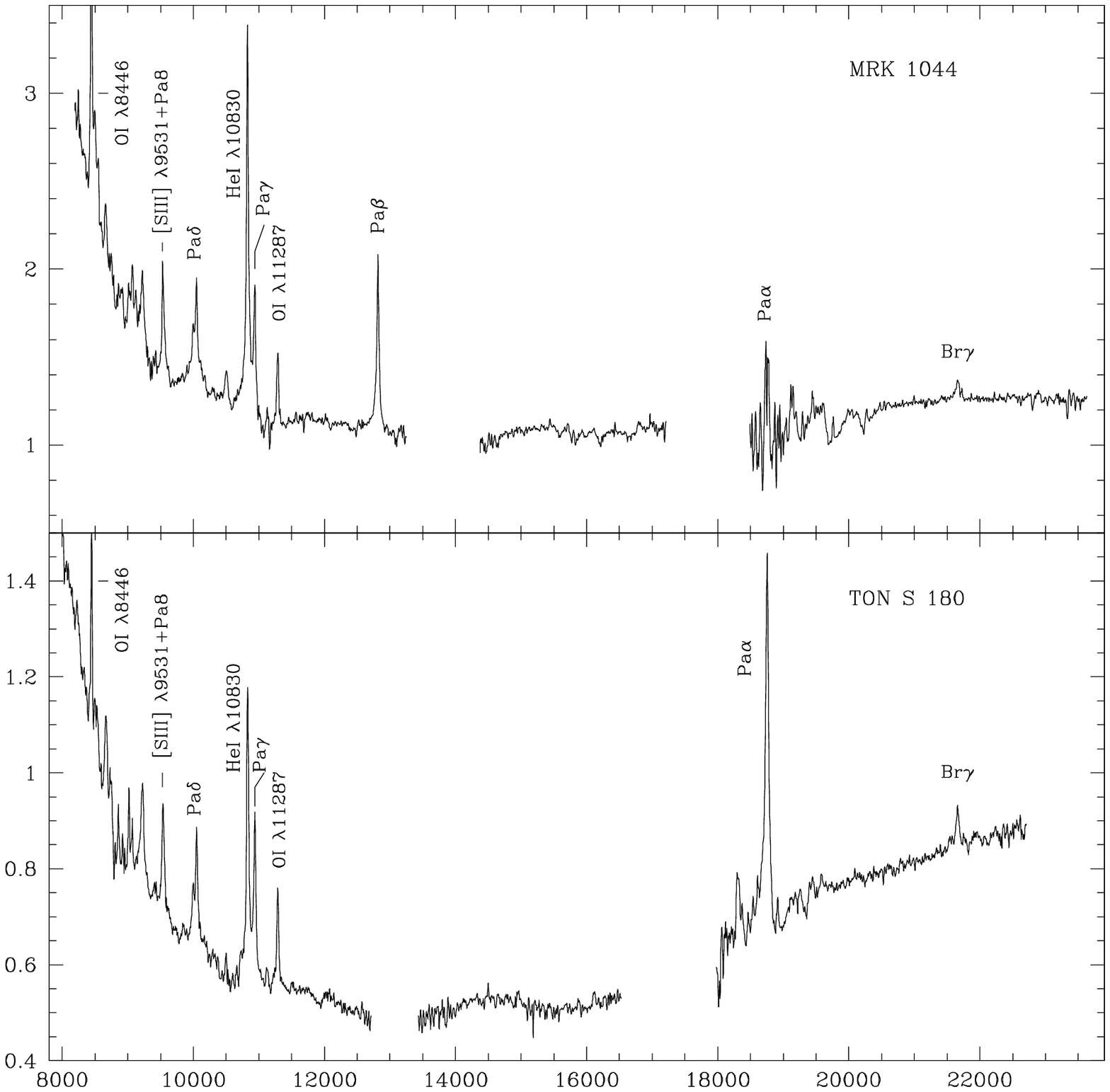}}
\figcaption{JHK spectrum of MRK\,1044 (upper panel) and TONS\,180 (botton
panel). Note the lack of significant coronal emission lines in these NLS1 objects.} \label{tonscor}

\centerline{\includegraphics[width=8.5cm]{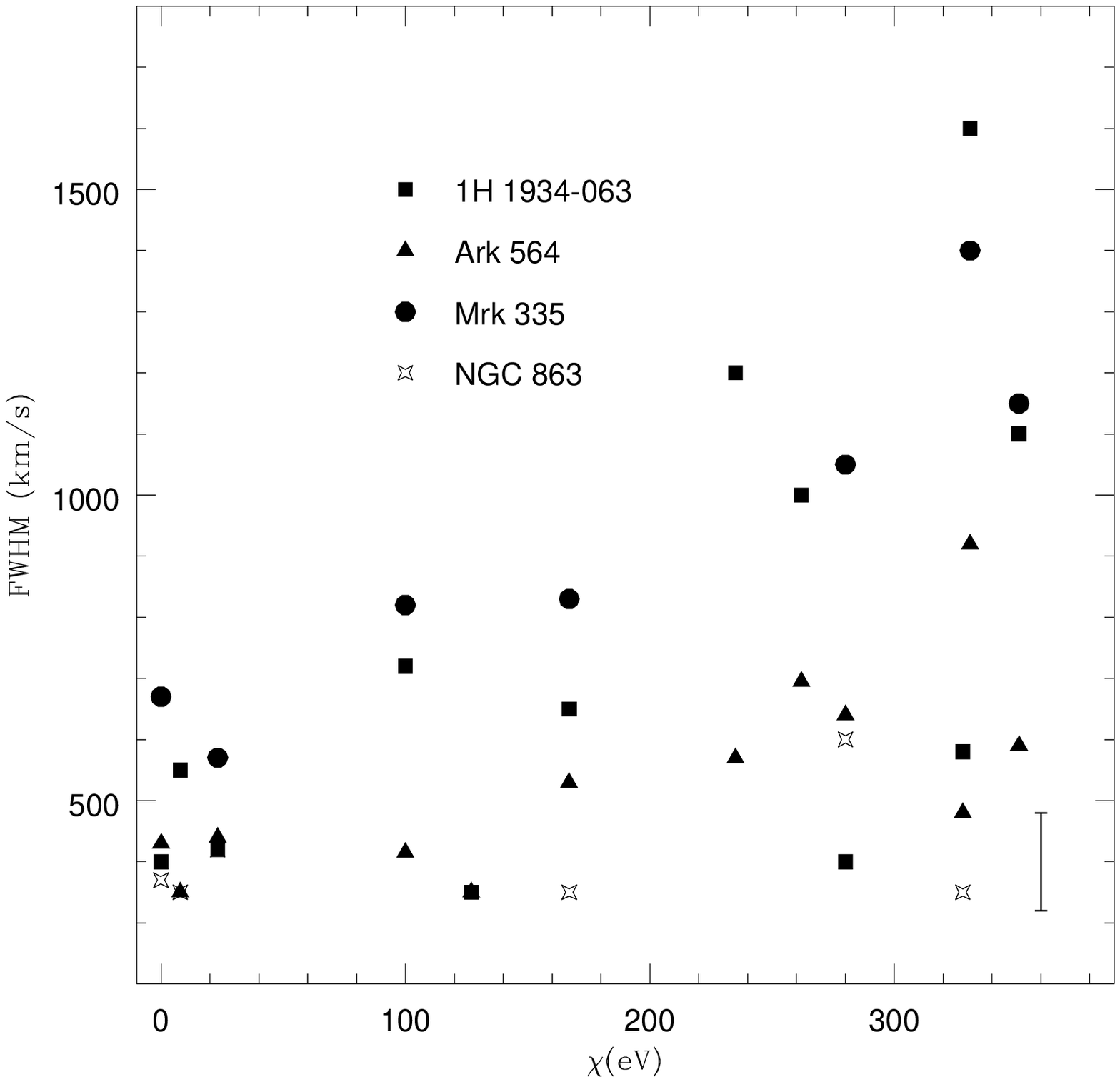}}
\figcaption{FWHM of the measured lines (already corrected for instrumental 
resolution) versus ionization potential. In NGC\,863, except for 
[\ion{S}{8}] 0.991 $\mu$m (IP=280 eV), the remaining forbidden emission were 
unresolved or detected as upper limits (see Section~\ref{kin} for 
further details). The magnitude of the largest FWHM uncertainty is
indicated by the error bar in the lower right corner.}
\label{ipvsfwhm}

\centerline{\includegraphics[width=8.5cm]{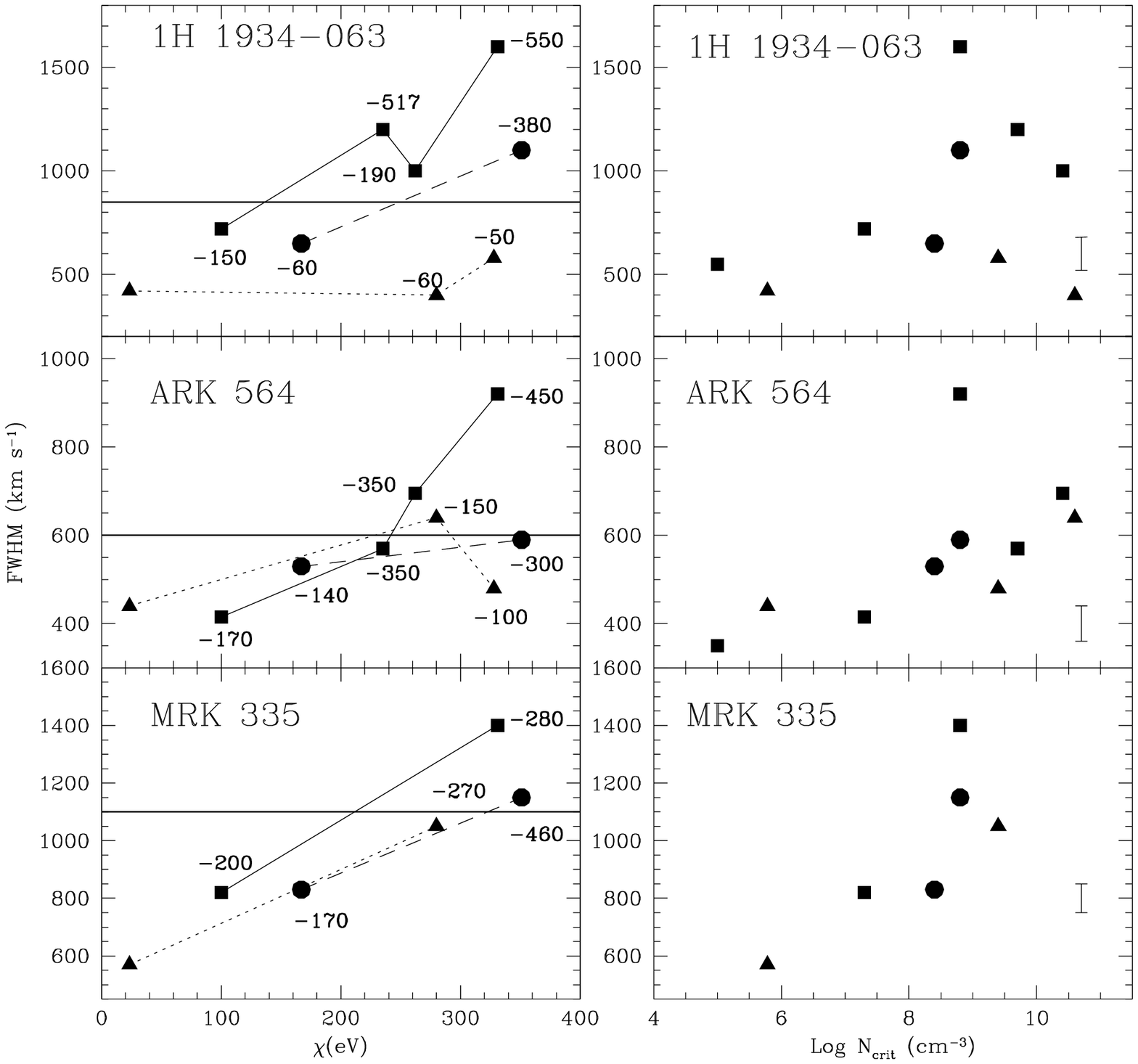}}
\figcaption{Observed FWHM versus ionization potential, $\chi$, (left panels) and FWHM
versus critical density, N$_{\rm crit}$, (right panels) for 1H\,1934-063 
(upper panel) ARK\,564 (middle panel) and MRK\,335 (lower panel) 
by chemical specie. Squares 
represents iron lines; triangles, sulfur lines and circles,
silicon lines. The thick line represents the FWHM of \ion{O}{1} 1.128 $\mu$m, 
a pure BLR line. The magnitude of the largest FWHM uncertainty is
indicated by the error bar in the right hand panels. The number beside each data point at the left panels is the shift of the line centroid regarding the systemic velocity of the galaxy. See text and Table~\ref{fwhm} for further details.} \label{ipvsfwhm1}

\centerline{\includegraphics[width=8.5cm]{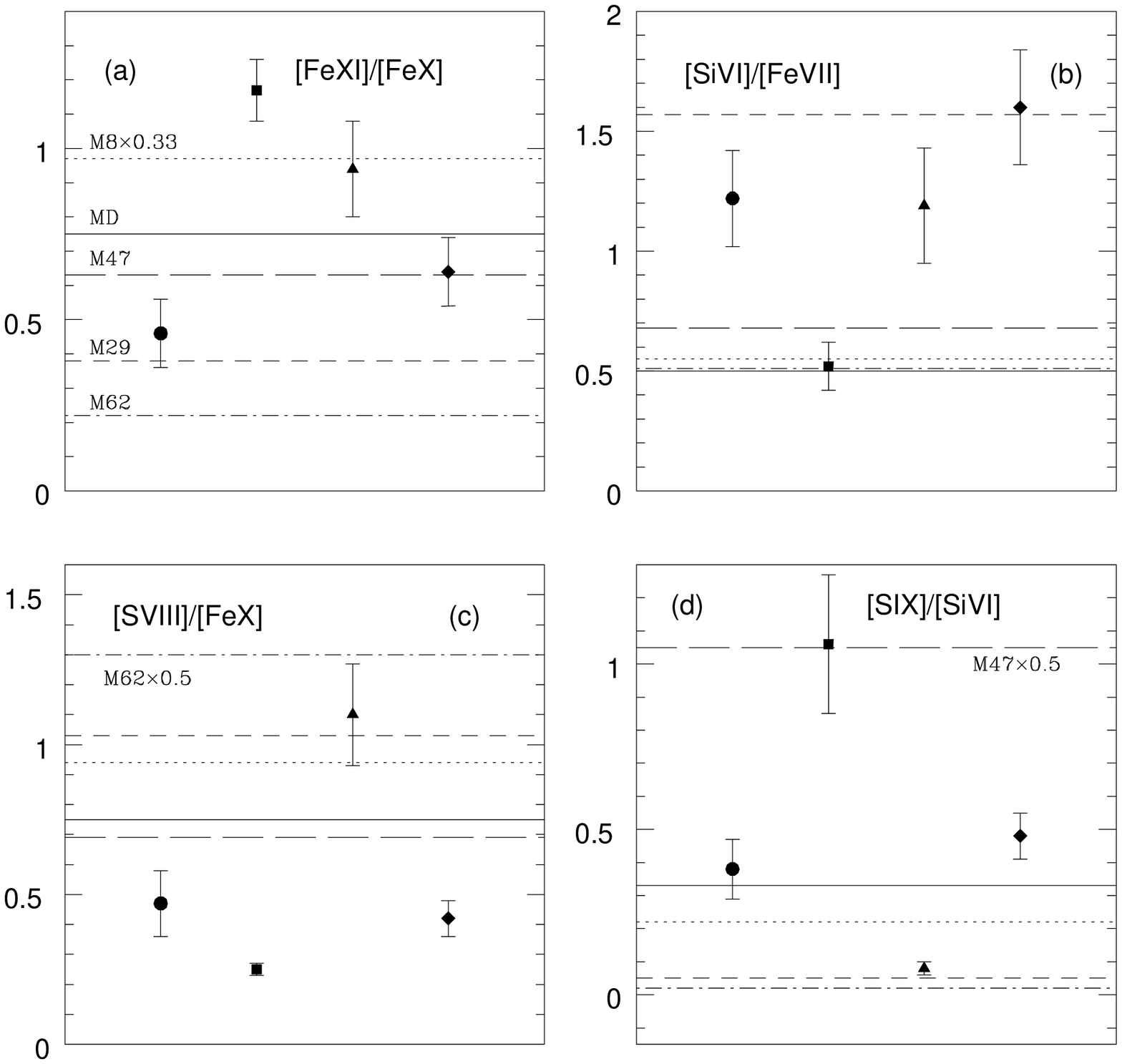}}
\figcaption{Comparison of model predictions with the observed, derredened
line ratios. 1H\,1934-063 is represented by the circle, ARK\,564 by the
square, NGC\,1068 by the triangle and Circinus by the diamond. The straight
line corresponds to model MD (composite model), the dotted line to model M8
($V_s$=100 \kms, $n_0$=100 cm$^{-3}$), the short-dashed
line to model M29 ($V_s$=200 \kms, $n_0$=200 cm$^{-3}$), the long-dashed 
line to model M47 ($V_s$=100 \kms, $n_0$=300 cm$^{-3}$) and the dot-dashed
line to model M62 ($V_s$=300 \kms, $n_0$=300 cm$^{-3}$). For visualization
purposes, M8 in panel (a), M62 in panel (c) and M47 in panel (d) were
multiplied by a scale factor (0.33, 0.5 and 0.5, respectively).} 
\label{obsvsmod}

  \begin{deluxetable}{lcccccc}
  \tablecaption{Sample characteristics. \label{tab1}}
  \tablewidth{0pt}  
  \tablehead{
  \colhead{Galaxy} & \colhead{z} & \colhead{M$_{\rm v}$\tablenotemark{a}} & 
  \colhead{A$_{\rm v}$\tablenotemark{b}} & \colhead{Type} & \colhead{N$_{\rm H}$\tablenotemark{c}} & \colhead{N$_{\rm H,Gal}$\tablenotemark{d}}}
  \startdata
  1H\,1934-063       & 0.01059 &  -19.04   & 0.972 & NLS1 & \nodata  & 18.8 \\
  Ark\,564          & 0.02468 &  -20.42   & 0.198 & NLS1 & 7.0$\pm$0.4\tablenotemark{1} & 6.4 \\
  Mrk\,335          & 0.02578 &  -21.32   & 0.118 & NLS1 & 3.9$\pm$0.2\tablenotemark{2} & 3.9$\pm$0.2 \\
  Mrk\,1044         & 0.01645 &  -18.84   & 0.113 & NLS1 & 4.2$\pm$0.4\tablenotemark{2} & 3.2 \\
  TON\,S\,180       & 0.06198 &  -22.57   & 0.047 & NLS1 & 1.15$\pm$0.45\tablenotemark{3} & 1.48 \\
  NGC\,863          & 0.02638 &  -21.27   & 0.124 & Seyfert 1 & 3.17$\pm$0.20\tablenotemark{2} & 2.72 \\
  \enddata
  \tablenotetext{a}{A value of H$_{\rm o}$ = 75 km s$^{-1}$ Mpc$^{-1}$ was assumed.}
  \tablenotetext{b}{Galactic extinction \citep{shl98}}
  \tablenotetext{c}{Hydrogen column density, in units of 10$^{20}$ cm$^{-2}$, derived from 
X-ray observations}
  \tablenotetext{d}{Galactic Hydrogen column density, in units of 10$^{20}$ cm$^{-2}$, derived
by \citet{dl90} from measuremnts of the 21-cm \ion{H}{1} line.}
\tablerefs{
(1) \citet{com01};
(2) \citet{pbr01};
(3) \citet{com98}.}
\end{deluxetable}

\begin{deluxetable}{lcccccc}
\tabletypesize{\footnotesize}
\tablecaption{Measured NIR narrow line fluxes for the galaxy sample\tablenotemark{1}. \label{fluxes}}
\tablewidth{0pt}  
\tablehead{
\colhead{Line} & \colhead{1H\,1934} & \colhead{ARK\,564} & \colhead{MRK\,335} & \colhead{NGC\,863} & \colhead{MRK\,1044} & \colhead{TONS\,180}}
\startdata
\lbrack \ion{Ca}{8} \rbrack \, 2.321$\mu$m & 3.94$\pm$1.10 &  3.12$\pm$1.00 & ...  & ...  & ...  & ... \\
Br$\gamma$\tablenotemark{b}\~\,2.165$\mu$m & 22.4$\pm$0.14& 10.50$\pm$0.6& 26.70$\pm$3.1& 9.34$\pm$2.60& 7.80$\pm$1.5& 5.30$\pm$1.00 \\
H$_{2}$ 2.121$\mu$m    &  1.07\tablenotemark{a}  &  1.14$\pm$0.40  & ...  & ... & ...   & ... \\
\lbrack \ion{Si}{6} \rbrack\, 1.963$\mu$m &  24.90$\pm$2.50&  7.20$\pm$1.30  & 10.61$\pm$2.0  & 2.14\tablenotemark{a} & ... & ... \\
H$_{2}$ 1.957$\mu$m  &  3.90$\pm$1.20 &  1.25$\pm$0.30 & ... & 3.70\tablenotemark{a} & ... & ... \\
Br$\delta$\tablenotemark{b}\~\,1.945$\mu$m&  5.04$\pm$1.70 &  8.04$\pm$0.60 & 20.27$\pm$4.0 & ...  & ... & ... \\
\lbrack \ion{Fe}{2} \rbrack\, 1.644$\mu$m&  14.33$\pm$3.50&  5.10$\pm$0.80& 1.85\tablenotemark{a} & ...        & ...  & ... \\
\lbrack \ion{Si}{10} \rbrack\, 1.4305$\mu$m&  20.00$\pm$5.40&  17.21$\pm$1.00& 7.45$\pm$3.00& 3.20$\pm$1.00 & ... & ... \\
Pa$\beta$\tablenotemark{b}\~\,1.282$\mu$m &  180.00$\pm$4.10&  98.50$\pm$1.00& 170.00$\pm$5.00& 58.00$\pm$3.00& 54.00$\pm$4.00& 54.8$\pm$5.00\tablenotemark{e} \\
\lbrack \ion{Fe}{2} \rbrack\,  1.257$\mu$m&  7.52$\pm$2.00&  3.67$\pm$0.60& ... & 1.00\tablenotemark{a} & ... & ... \\
\lbrack \ion{S}{9} \rbrack\, 1.252$\mu$m&  9.00$\pm$2.00&  7.51$\pm$0.64 & 12.60$\pm$2.5\tablenotemark{d} & 1.10$\pm$0.35 & ... & ... \\
\lbrack \ion{P}{2} \rbrack\, 1.188$\mu$m&  ... & 4.47$\pm$0.90 & ... & ... & ... &  \\
\lbrack \ion{Ni}{2} \rbrack\, 1.191$\mu$m & ... & 1.33$\pm$0.40 & ... & ... & ... &  \\
\lbrack \ion{Fe}{13} \rbrack\, 1.074$\mu$m& 29.20$\pm$4.00&  10.74$\pm$2.00 & 18.33$\pm$3.80 & ... & ... & 1.23$\pm$0.4  \\
\lbrack \ion{S}{2} \rbrack\, 1.030$\mu$m\tablenotemark{c} & 12.18$\pm$3.41&  5.57$\pm$1.30& 1.83\tablenotemark{a} & 2.31$\pm$0.52& 1.85$\pm$0.30& ... \\
\lbrack \ion{S}{8} \rbrack\, 0.991$\mu$m&  12.30$\pm$2.80&  6.54$\pm$0.40& 3.76$\pm$1.00& 1.00\tablenotemark{a}  & ... & ... \\
\lbrack \ion{Ca}{1} \rbrack\, 0.985$\mu$m&  3.67$\pm$1.00&  0.88$\pm$0.25& 1.00\tablenotemark{a}& 1.45$\pm$0.6  & ...  & 0.9\tablenotemark{a} \\
\lbrack \ion{S}{3} \rbrack\, 0.953$\mu$m&  165.00$\pm$3.80& 31.05$\pm$1.00& 22.5$\pm$2.30 & 15.20$\pm$0.60& 20.90$\pm$0.80& 2.43\tablenotemark{a} \\
\enddata
\tablenotetext{1}{In units of 10$^{-15}$ erg cm$^{-2}$ s$^{-1}$.} 
\tablenotetext{a}{Upper limit.} 
\tablenotetext{b}{Sum of broad and narrow fluxes.}
\tablenotetext{c}{Sum of the fluxes of the [\ion{S}{2}] lines located at 1.028$\mu$m and 1.032$\mu$m.}
\tablenotetext{d}{Strongly blended with \ion{He}{1} $\lambda$1.253.}
\tablenotetext{e}{Flux of Pa$\alpha$.}
\end{deluxetable}

\begin{deluxetable}{lccccccc}
\tabletypesize{\scriptsize}
\tablecaption{Line ratios between the most important forbidden lines. \label{corrat}}
\tablewidth{0pt}  
\tablehead{
\colhead{Ratio} & \colhead{1H\,1934} & \colhead{ARK\,564} & \colhead{MRK\,335} & \colhead{NGC\,863} & \colhead{NGC\,1068} & \colhead{Circinus\tablenotemark{c}} & \colhead{NGC\,4151\tablenotemark{d}}}
\startdata
\lbrack \ion{Ca}{8} \rbrack/H$_{2}$ 2.121$\mu$m  & 3.68\tablenotemark{2} & 2.74$\pm$1.30 & ... & ... & ...  & 0.88$\pm$0.13 & ... \\
\lbrack \ion{Si}{6} \rbrack/H$_{2}$ 1.957$\mu$m& 6.38$\pm$2.06 & 5.76$\pm$1.73 & ... & 0.57\tablenotemark{2}& 5.00$\pm$1.70\tablenotemark{b} & 1.38$\pm$0.20 & 5.70$\pm$1.93 \\
\lbrack \ion{Fe}{2} \rbrack/\lbrack \ion{Si}{6} \rbrack\, & 0.58$\pm$0.15 & 0.71$\pm$0.17 & 0.17\tablenotemark{1}& ...& ...& 0.85$\pm$0.13 & 1.10$\pm$0.10 \\
\lbrack \ion{Si}{10} \rbrack/\lbrack \ion{Fe}{2} \rbrack\, 1.257$\mu$m& 2.66$\pm$1.01 & 4.63$\pm$0.80 & ... & 3.2\tablenotemark{2} & ... & ...& ... \\           
\lbrack \ion{S}{9} \rbrack/\lbrack \ion{Fe}{2} \rbrack\, 1.257$\mu$m & 1.20$\pm$0.40 & 2.04$\pm$0.38& ...& 1.1\tablenotemark{2}& 0.65$\pm$0.10\tablenotemark{a}& 0.34$\pm$0.05 & 0.17$\pm$0.04 \\
\lbrack \ion{S}{2} \rbrack/\lbrack \ion{S}{8} \rbrack & 0.99$\pm$0.35& 0.65$\pm$0.15& 0.49\tablenotemark{1}& 2.31\tablenotemark{2}& 3.09\tablenotemark{a,1} & ... & ... \\
\lbrack \ion{S}{8} \rbrack/\lbrack \ion{Ca}{1} \rbrack & 3.35$\pm$1.18& 9.80$\pm$2.80& 3.76\tablenotemark{2}& 0.69\tablenotemark{1}& 0.77$\pm$0.12\tablenotemark{a}& 2.00$\pm$0.30  & ... \\
\lbrack \ion{S}{3} \rbrack/\lbrack \ion{S}{8} \rbrack & 13.41$\pm$3.07& 4.74$\pm$0.45& 5.98$\pm$1.70& 15.20\tablenotemark{2}& 21.82$\pm$3.27\tablenotemark{a}& 14.00$\pm$2.1 & 27.30$\pm$4.60 \\
\enddata
\tablenotetext{1}{Upper limit.}
\tablenotetext{2}{Lower limit.}
\tablerefs{
(a) \citet{oli01};
(b) \citet{mar94};
(c) \citet{oli94};
(d) \citet{tho95}.}

\end{deluxetable}   

\begin{deluxetable}{lcccc}
\tablecaption{Optical emission line fluxes\tablenotemark{1} \, for the galaxy sample.\label{opfluxes}}
\tablewidth{0pt}
\tablehead{
\colhead{} & \colhead{[\ion{Fe}{7}]} & \colhead{[\ion{Fe}{10}]} & \colhead{[\ion{Fe}{11}]} \\
\colhead{Galaxy} & \colhead{\lb6087} & \colhead{\lb6374} & \colhead{\lb7892} & \colhead{E(B-V)}}
\startdata
1H\,1934 & 16.20$\pm$2.20 &  23.20$\pm$2.30 & 11.77$\pm$2.50 & 0.10 \\
ARK\,564\tablenotemark{a} & 13.00$\pm$0.67 &  25.46$\pm$1.51 & 30.40$\pm$1.42 & 0.03 \\
MRK\,335     & 4.52$\pm$0.57  &  2.14$\pm$0.51  &  ...         & 0.18 \\
MRK\,1044    & 4.36$\pm$1.00  &  2.83\tablenotemark{b}     &  ...         & 0.05 \\
TON\,S\,180  & 1.89$\pm$0.60  &   ...         &  ...         & 0.00 \\
\enddata
\tablenotetext{1}{In units of 10$^{-15}$ erg cm$^{-2}$ s$^{-1}$.}
\tablenotetext{a}{Data taken from \citet{eaw97}.}
\tablenotetext{b}{Upper limit.}
\end{deluxetable}

\begin{deluxetable}{lccccccccccccc}
\tabletypesize{\scriptsize}
\tablecaption{FWHM\tablenotemark{1} \, of coronal and forbidden low ionization lines. \label{fwhm}}
\tablewidth{0pt}
\tablehead{
&  & \multicolumn{2}{c}{1H\,1934} & & \multicolumn{2}{c}{ARK\,564} & & \multicolumn{2}{c}{MRK\,335} & & \multicolumn{2}{c}{NGC\,863} & \colhead{Log N$_{\rm crit}$}\\
\cline{3-4} \cline{6-7} \cline{9-10} \cline{12-13}
\colhead{Line} & \colhead{$\chi$ (eV)} & \colhead{FWHM} & \colhead{$\Delta$V} & & \colhead{FWHM} & \colhead{$\Delta$V} & & \colhead{FWHM} & \colhead{$\Delta$V} & & \colhead{FWHM} &
\colhead{$\Delta$V} & \colhead{(cm$^{-3}$)}}

\startdata
\lbrack \ion{Fe}{7} \rbrack & 100  & 720 & -150 & & 420\tablenotemark{c}& -170\tablenotemark{c}& & 820& -200& & ... & ... & 7.3 \\
\lbrack \ion{Fe}{10} \rbrack & 235  & 1200& -517 & & 570\tablenotemark{c}& -350\tablenotemark{c}& & ...& ...&  & ... & ... & 9.7 \\
\lbrack \ion{Fe}{11} \rbrack & 262  & 1000& -190 & & 695\tablenotemark{c}& -350\tablenotemark{c}& & ...& ...&  & ... & ... & 10.4 \\
\lbrack \ion{S}{3} \rbrack & 23.3 & 420 &  ... & & 440 &  ...  & & 570 & ...  &    & 420 & ... & 5.8 \\
\lbrack \ion{C}{1} \rbrack & 0.0  & 400\tablenotemark{a}&...& & 430 & +60   & & 670 & +120 &    & 370 & +60 & 4.3 \\
\lbrack \ion{S}{8} \rbrack & 280  & 400 & -60  & & 890 & -150  & & 1050& -270 &    & 600 & -90 & 10.6 \\
\lbrack \ion{Fe}{13} \rbrack & 331  & 1600& -550 & & 920 & -450  & & 1400& -280 &    &  ...& ... & 8.8 \\
\lbrack \ion{S}{9} \rbrack & 328  & 580 & -50  & & 480 & -100  & & ... & ...  &    & 350 & -50 & 9.4 \\
\lbrack \ion{Fe}{2} \rbrack & 7.9  & 550 & +70  & & 350 &  ...  & & ... & ...  &    & 350 & -100 & 5.0 \\
\lbrack \ion{Si}{10} \rbrack & 351  & 1100& -380 & & 590 & -300  & & 1150& -460 &    & ... & ...  & 8.8 \\
\lbrack \ion{Si}{6} \rbrack & 167  & 650 & -60  & & 530 & -140  & & 830 & -170 &    & 350 & ... & 8.8 \\
\lbrack \ion{Ca}{8} \rbrack & 127  & 350 & -150 & & 350 & ...   & & ... & ...  &    & ... & ...&  7.9 \\
Pa$\beta^{b}$ &   0.0  & 1800& ...  & & 1800& ...   & & 2050& ...  &    & 2720& ... & ... \\
\ion{O}{1}~1.1287$\mu$m & 0.0 & 850 & ... & & 600 & ... & & 1100& ... &  & 2000& ... & ... \\
\enddata
\tablenotetext{1}{In km s$^{-1}$, already corrected for instrumental FWHM. The 
columns labeled $\Delta$V correspond to the shift of the line relative to the 
systemic velocity of the galaxy. A plus sign (+) represents a redshift while 
a negative sign (-) a blueshift.}
\tablenotetext{a}{Upper limit.}
\tablenotetext{b}{FWHM of the broad component.}
\tablenotetext{c}{Taken from \citet{eaw97}.}
\end{deluxetable}

\begin{deluxetable}{lcccccccccc}
\tabletypesize{\footnotesize}
\tablecaption{Reddening corrected ratios among optical and NIR coronal lines and model predictions. \label{opnir}}
\tablewidth{0pt}
\tablehead{
\colhead{Ratio} & \colhead{1H\,1934} & \colhead{ARK\,564} & \colhead{NGC\,1068\tablenotemark{a}} & \colhead{Circinus\tablenotemark{b}} & \colhead{MD\tablenotemark{c}} & \colhead{M8\tablenotemark{d}} & \colhead{M29\tablenotemark{d}} & \colhead{M47\tablenotemark{d}} & \colhead{M62\tablenotemark{d}} & \colhead{PH2\tablenotemark{e}}} 
\startdata
\lbrack \ion{Fe}{10} \rbrack/\lbrack \ion{Fe}{7} \rbrack & 1.41$\pm$0.23 & 1.95$\pm$0.15  & 0.16$\pm$0.02 & 1.89$\pm$0.32 & 0.67 & 0.88 & 0.48 & 6.98 & 0.06 & 0.41 \\
\lbrack \ion{Si}{6} \rbrack/\lbrack \ion{Fe}{7} \rbrack & 1.22$\pm$0.20 & 0.52$\pm$0.10 & 1.19$\pm$0.24 & 1.60$\pm$0.24 & 0.50 & 0.55 & 1.57 & 0.68 & 0.51 & 0.81 \\
\lbrack \ion{Fe}{11} \rbrack/\lbrack \ion{Fe}{10} \rbrack & 0.46$\pm$0.10 & 1.17$\pm$0.09 & 0.94$\pm$0.14 & 0.64$\pm$0.10 & 0.75 & 2.91 & 0.38 & 0.63 & 0.22 & 0.54 \\
\lbrack \ion{Fe}{13} \rbrack/\lbrack \ion{Fe}{10} \rbrack & 1.01$\pm$0.17 & 0.40$\pm$0.08  & ... & ... & ... & ... & ... & ... & ... & ... \\
\lbrack \ion{S}{8} \rbrack/\lbrack \ion{Fe}{10} \rbrack & 0.47$\pm$0.11 & 0.25$\pm$0.02 & 1.1$\pm$0.17  & 0.42$\pm$0.06 & 0.75 & 0.94 & 1.03 & 0.69 & 2.60 & 0.34 \\
\lbrack \ion{S}{9} \rbrack/\lbrack \ion{Si}{6} \rbrack &  0.38$\pm$0.09 & 1.06$\pm$0.21 & 0.08$\pm$0.02 & 0.48$\pm$0.07 & 0.33 & 0.22 & 0.05 & 2.1 & 0.02 & 0.06 \\ 
\enddata
\tablenotetext{a}{Taken from \citet{mar96}.}
\tablenotetext{b}{Taken from \citet{oli94}.}
\tablenotetext{c}{Model label MD from \citet{cpv98}.}
\tablenotetext{d}{M8, M29, M47 and M62 stand for models numbered 8, 29, 47 and 62 in \citet{cv01}.}
\tablenotetext{e}{Line ratios predicted by a pure photoionization model assuming the standard AGN continuum, taken from \citet{oli94}} 
\end{deluxetable}

\end{document}